\documentclass[%
 aip,
 amsmath,amssymb,
 reprint,%
]{revtex4-1}

\usepackage{graphicx}
\usepackage{dcolumn}
\usepackage{bm}

\usepackage[utf8]{inputenc}
\usepackage[T1]{fontenc}
\usepackage{mathptmx}

\begin{document}

\preprint{AIP/123-QED}

\title{Mobile obstacles accelerates and inhibits the bundle formation in two-patch colloidal particle}

\author{Isha Malhotra and Sujin B. Babu}
  \altaffiliation[Also at ]{Department of physics, Indian Institute of Technology, Delhi-110016}

\date{\today}

\begin{abstract}
Aggregation of protein into bundles is responsible for many neurodegenerative diseases. In this work, we show how two-patch colloidal particles self assemble into chains and a sudden transition to bundles takes place by tuning the patch size and solvent condition. We study the kinetics of formation of chains, bundles and network like structures using Patchy Brownian cluster dynamics. We also analyse the ways to inhibit and accelerate the formation of these bundles. We show that in presence of inert immobile obstacles, the kinetics of formation of bundles slows down. Whereas, in presence of mobile aggregating particles which exhibit inter-particle attraction and intra-particle repulsion, the kinetics of bundle formation accelerates slightly. We also show that if we introduce mobile obstacles which exhibit intra-particle attraction and inter-particle hard sphere repulsion, the kinetics of formation of bundles is inhibited. This is similar to the inhibitory effect of peptide P4 on the formation of insulin fibres. We are providing a model of mobile obstacles undergoing directional interactions to inhibit the formation of bundles.
\end{abstract}

\maketitle

\section{\label{sec:level1}Introduction}

Self-assembly of monomers into one-dimensional aggregates is found in many systems such as carbon nanotubes\cite{dresselhaus1998physical,shulaker2013carbon}, DNA wires\cite{berlin2000dna,de2011self}, tubular surfactant micelles\cite{meng2012tunable,daful2012model}, amyloid  fibres\cite{scheibel2003conducting,xue2008systematic}. The formation of amyloid fibers is responsible for many neurodegenerative diseases such as type II diabetes, Alzheimer's disease and Parkinson's disease\cite{huang2012alzheimer,soto2003unfolding,hardy2002amyloid}.  In order to achieve control over the formation of these fibers, it is important to understand the underlying mechanism of their formation. Direct control over fibrillization process may provide effective therapeutical strategies to treat these neurodegenerative diseases. Various approaches are proposed to inhibit fibrillation such as adding small molecules like flavonoids\cite{wang2015mechanisms}, vitamins\cite{alam2016vitamin}, metal chelators\cite{liu2017iminodiacetic}, nanoparticles\cite{barros2018gold} that can interfere with the mechanism of formation of these fibers. One of the ways to mimic protein aggregation is by using patchy particles\cite{kraft2012surface, duguet2016patchy}. It has already been shown that patchy particle model developed by Kern and Frenkel\cite{kern2003fluid} reproduce many type of protein crystals\cite{roberts2014role,chen2011directed,romano2011two,noya2010stability,noya2007phase,noya2008computing,doye2007controlling} and ordered structures\cite{romano2012phase,romano2011crystallization,vissers2013predicting,romano2011colloidal,romano2012patterning}. Huisman \textit{et al.}\cite{huisman2008phase} have shown that the formation of bundles is similar to sublimation transition of polymers. For the case of one-patch colloidal particle, small clusters spontaneously reorganize into long straight tubes at specific temperatures and densities\cite{munao2013cluster,preisler2013phase,vissers2014cooperative}.   Patchy particles are also considered to be building blocks of various supracolloidal helices\cite{zou2016supracolloidal,morgan2013designing,guo2014predictive}. Two patch model along with isotropic interaction reproduced the chain to bundle transition usually seen in Lysozyme protein\cite{malhotra2018aggregation,woodard2014gel}. They also showed the coexistence of thermodynamically favored bundles along with equilibrium crystals\cite{malhotra2019phase}.

In the present work we introduce a method to construct bundles through self-assembly of patchy particles under different solvent conditions with tunable patch size using Patchy Brownian cluster Dynamics\cite{prabhu2014brownian}. For the case of isotropic square well, this simulation technique gives exactly the same results as Event driven Brownian dynamics $(EDBD)$\cite{babu2006phase}. This technique has also been modified to study the binary system of colloidal particles\cite{shireen2017lattice,shireen2018cage}. This algorithm when applied to single polymer chain gave correct dynamic and static properties\cite{prabhu2014brownian}. 

By tuning both patch size and solvent condition, we are able to find the sharp transition point of chains to bundles which is responsible for many neurodegenerative diseases. We also investigate the effect of immobile and mobile obstacles on the formation of bundles. Experimentally many inhibitors are proposed to attenuate the formation of bundles \cite{feng2008small,lendel2010detergent}. These inhibitors interact in a non-specific manner with bundle forming particles. In this work, we are identifying the specific interactions that can inhibit bundle formation even in presence of very low fraction of obstacles.  

This paper is arranged in the following way. In section 2, we introduce our model and simulation technique and explain how we implement the interaction between different kinds of particles. In section 3, the results of our simulations are discussed. We show the effect of immobile and mobile obstacles on the formation of bundles. Immobile obstacles ($I$ particles) decelerates the formation of bundles. The presence of mobile obstacles ($M$ particles) slightly accelerates the formation of bundles if they exhibit inter-particle attraction and interact via hard sphere repulsion with bundle forming $B$ particles. Whereas mobile obstacles inhibit the formation of bundles, if they exhibit inter-particle hard sphere repulsion and aggregate with bundle forming $B$ particles. In section 4, main findings of our work are concluded. 
\section{\label{sec:model}Model and simulation technique}
\begin{figure}
\includegraphics[height=2.5cm,width=8.5cm]{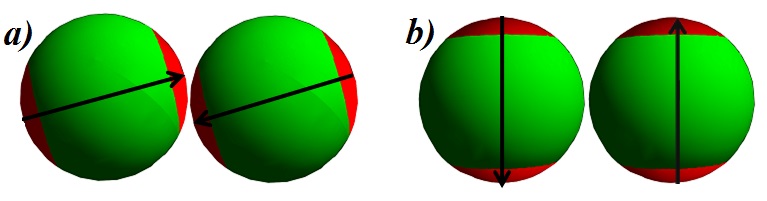}
  \caption{Red region represents the patchy part of the sphere and green region represents the non patchy part of the sphere. (a).  Particles are in interaction range and patches are facing each other and hence forms irreversible $P$ bond. (b). Particles are in interaction range and as patches are not facing each other, they form $NPI$ bond with a probability $1-exp(-u_{i})$.}
    \label{PNPI}
    \end{figure}

In the present study, we start with $N$ randomly distributed two-patch colloidal particles ($B$ particles) each of diameter $\sigma=1$ in a three dimensional cubic box of length $L=50$ with periodic boundary condition. The volume fraction is defined as $\phi=(\pi/6) N/L^3$. In the present work, we have kept $\phi=0.02$ as it has been shown that typical volume fraction of amyloid gel of lysozyme protein is $0.02$\cite{woodard2014gel}. This corresponds to $N$=$4774$ $B$ particles. Each $B$ particle has two oppositely located patches which are defined by a unit patch vector $\hat{\bf v}_{i}$. To simulate interparticle anisotropic interaction, we employ the model developed by Kern and Frenkel\cite{kern2003fluid}.  We couple reversible isotropic potential with irreversible anisotropic potential, hence total potential $U (r_{i,j}, v_{i}, v_{j})$ is defined as: 
\begin{equation}
 U (\mathbf{r}_{i,j},  {\bf v}_{i} , {\bf v}_{j})  = 
 \begin{cases}
   \infty\hspace{2.2cm} r_{i,j}\leq \sigma 
   \\
   -(u_{i} + u_{a})\quad \sigma < r_{i,j}\leq\sigma(1+\epsilon)
\\
0\hspace{2.3cm}  r_{i,j}>\sigma(1+\epsilon)
\end{cases}
\label{e.1}
 \end{equation}

 where $\epsilon=0.1$ is the interaction range which is kept identical for both isotropic and anisotropic interactions, $r_{i,j}$ is the inter particle distance between their centres of mass. $u_{i}$ and $u_{a}$ are depths of the square well for isotropic interactions and anisotropic interactions respectively. $u_{a}$ is non zero only if $ {\textbf{\^r}}_{i,j}.{\textbf{\^v}}_{i}\ \text{and}\         {\textbf{\^r}}_{j,i}.{\textbf{\^v}}_{j}>\text{cos}\ $ $\omega$. $\omega$ defines the patch size and it's a tunable parameter. In the present work we have considered $\omega$ ranging from $12^{\circ}$ to $40^{\circ}$ for simulating $B$ particles. We are allowing multiple bonds to form via patches. To investigate the formation of bundles in presence of obstacles, we randomly distribute obstacles along with $B$ particles in the box at the start of simulation. The fraction of these obstacles $(C_0)$ is defined as $C_0=\frac{N_0}{N}$, $N_0$ is the number of $I$ particles (if obstacles are immobile) and $M$ particles (if obstacles are mobile). $C_0$=1 means that we have equal number of $B$ particles and obstacles. Diameter of each obstacle is also kept as unity. During movement step, $2N_{tot}$ ($N_{tot}=N+N_O$, if obstacles are mobile else $N_{tot}=N$) times a particle is randomly selected. It is either translated or rotated with equal probability only if this movement step doesn't lead to overlap or breakage of bond with any of the neighboring particles. In order to ensure correct diffusional behaviour, we have fixed our rotational step size $s_R=0.018$ and translational step size $s_T=0.0132$ \cite{prabhu2014brownian}. The cumulative effect of this step is the diffusion of center of mass of clusters. After this step, cluster construction step is carried out, where if particles satisfy the condition to form a $Patchy$ bond, see Fig. \ref{PNPI}a then an irreversible bond is formed. The collection of these bonded particles are said to form $P$ type cluster. When particles are in each other's interaction range but do not satisfy the conditions to form a $Patchy $ bond, then a $Non$ $patchy$ $isotropic(NPI)$ (see Fig. \ref{PNPI}b) bond is formed with a probability $\alpha_{i}$ and an existing bond is broken with a probability $\beta{i}$, such that $\frac{\alpha_{i}}{\alpha_{i}+\beta_{i}}=1-\exp(-u_{i})$\cite{prabhu2014brownian}. $u_{i}$ is given in terms of $K_BT$ and $K_BT$ is kept as unity. If a cluster extends from one end of the box to another end in any direction, we call it as a percolated cluster. For the isotropic interaction, we define second virial coefficient $B_2=B_{rep}-B_{att}$, where $B_{rep}=4$ is the repulsive part of the interaction due to hard sphere interaction between the spheres. $B_{att}$ is the attractive part of the potential which is defined in terms of interaction range and isotropic potential depth as $B_{att}=4.[exp(-u_{i}).[(1+\epsilon^2)-1]]$ \cite{babu2006phase}. In this study we have used range of $B_{att}$ values from 4 to 12 and quality of solvent deteriorates as we move from lower $B_{att}$ value towards higher $B_{att}$ value. In this study we have reported time in reduced units $t/t_0$, where $t_0$ is defined as the time taken by a particle to travel a distance equal to it's own diameter \cite{prabhu2014brownian}. 

In the present study three systems are considered for obstacles.
\begin{enumerate}
\item Immobile obstacles
\\In this system, we investigate the aggregation of particles in presence of $C_0$ fraction of obstacles which are frozen and do not diffuse throughout the simulation. They interact via hard sphere repulsion with $B$ particles. 
\item Mobile obstacles with inter-particle attraction and  hard sphere intra-particle repulsion
\\Here, we investigate the aggregation of $B$ particles in presence of $C_0$ fraction of obstacles. These obstacles interact with irreversible isotropic square well interaction (when particles are in each other's interaction range, they form an irreversible bond) with each other and there is only hard sphere repulsion with $B$ particles.
\item Mobile patchy obstacles with intra-particle attraction and inter-particle hard sphere repulsion
\\We take $C_0$ fraction of patchy obstacles with patch size $\omega=90^{\circ}$ (isotropic square well) along with $B$ particles to investigate the formation of bundles. These obstacles interact with $B$ particles via potential given by Eq. \ref{e.1}, whereas they interact via hard sphere repulsion with each other.
\end{enumerate}

\section{\label{sec:result}Results}
\begin{figure}
\includegraphics[height=10cm,width=8.5cm]{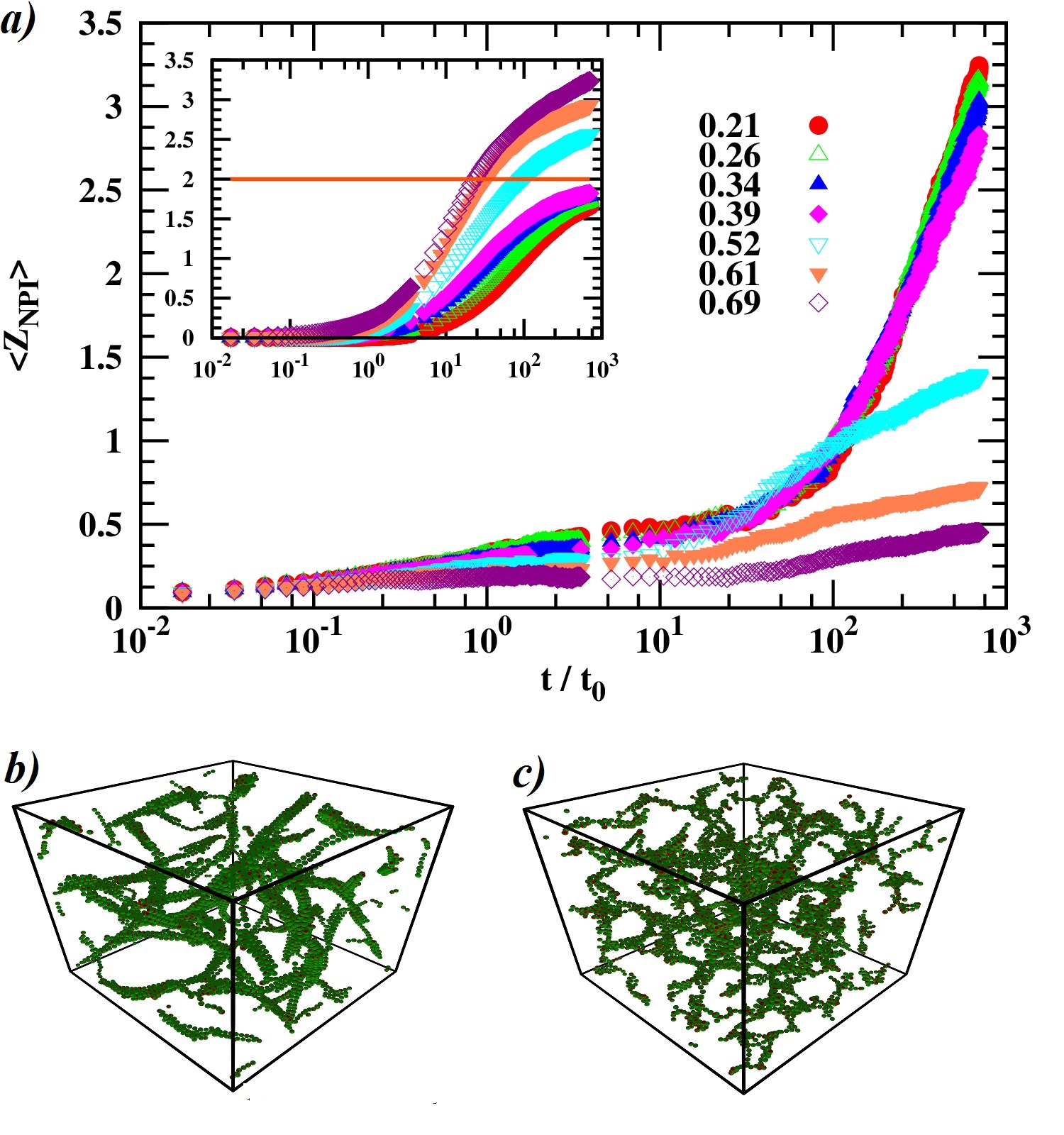}
  \caption{(a). The average bonded neighbors for $NPI$ type cluster $<$$Z_{NPI}$$>$ is plotted with respect to reduced time at different $\omega$ values as indicated in the figure at $B_{att}=12$. The inset shows average bonded neighbors for $P$ type cluster $<$$Z_P$$>$. Solid line indicates that $<$$Z_P$$>$$<$$2$ for $\omega\leq22.5^{\circ}$(b). Snapshot of system formed at $\omega=22.5^{\circ}$ is shown at $t/t_0=1800$, where we can observe bundles.(c). Snapshot of system formed at $\omega=30^{\circ}$ is shown, where we can observe several flower like structures and it is locally denser than the system formed at $\omega=22.5$.
    }
  \label{B12NPI}
\end{figure}

In order to understand the kinetics of aggregation of system, we follow the average number of neighbors $<$$Z_P$$>$ and $<$$Z_{NPI}$$>$ as a function of reduced time. In Fig. \ref{B12NPI}, we have plotted the average number of neighbors for $NPI$ clusters ($<$$Z_{NPI}$$>$) at $B_{att}=12$ at different $\omega$ values as indicated in the figure.  In inset, we have plotted $<$$Z_P$$>$ as a function of reduced time and we observe that $<$$Z_P$$>$ $\sim2$ for $\omega\leq22.5^{\circ}$. It indicates that on average for $\omega\leq22.5^{\circ}$, one bond per patch is possible and system consists of chains formed via irreversible anisotropic interaction. We observe that $<$$Z_{NPI}$$>$ value for $\omega\leq22.5^{\circ}$ after a plateau takes a sudden upturn at $t/t_0\geq100$. This upturn is similar to nucleation and growth mechanism\cite{babu2006phase} and it indicates chain to bundle transition \cite{malhotra2018aggregation}. The snapshot of system formed at $\omega=22.5^{\circ}$ is shown in Fig. \ref{B12NPI}b where the presence of bundles is clearly visible. For $\omega\geq30^{\circ}$, $<$$Z_P$$>>2$ indicating that on average each patch is able to form multiple bonds per patch via irreversible anisotropic interaction. $<$$Z_{NPI}$$>$ for $\omega\geq30^{\circ}$ is less than $<$$Z_P$$>$ due to increased patch size and irreversibility of patchy bond. This indicates that irreversible anisotropic interaction is playing a major role in aggregation of system. Irreversible aggregation always leads to branching and results in the formation of network like structure as shown in the snapshot of system formed at $\omega=30^{\circ}$ in Fig. \ref{B12NPI}c. This network is locally denser than system formed at $\omega=22.5^{\circ}$ due to increased patch size. We also observe that system percolates faster when patch size is smaller. This is due to the fact that system formed at higher patch size is locally denser and hence percolates slower. It indicates that in order to observe bundles at $B_{att}=12$, patches should form single bond\cite{malhotra2018aggregation}. 

\begin{figure}
\includegraphics[height=10cm,width=8.5cm]{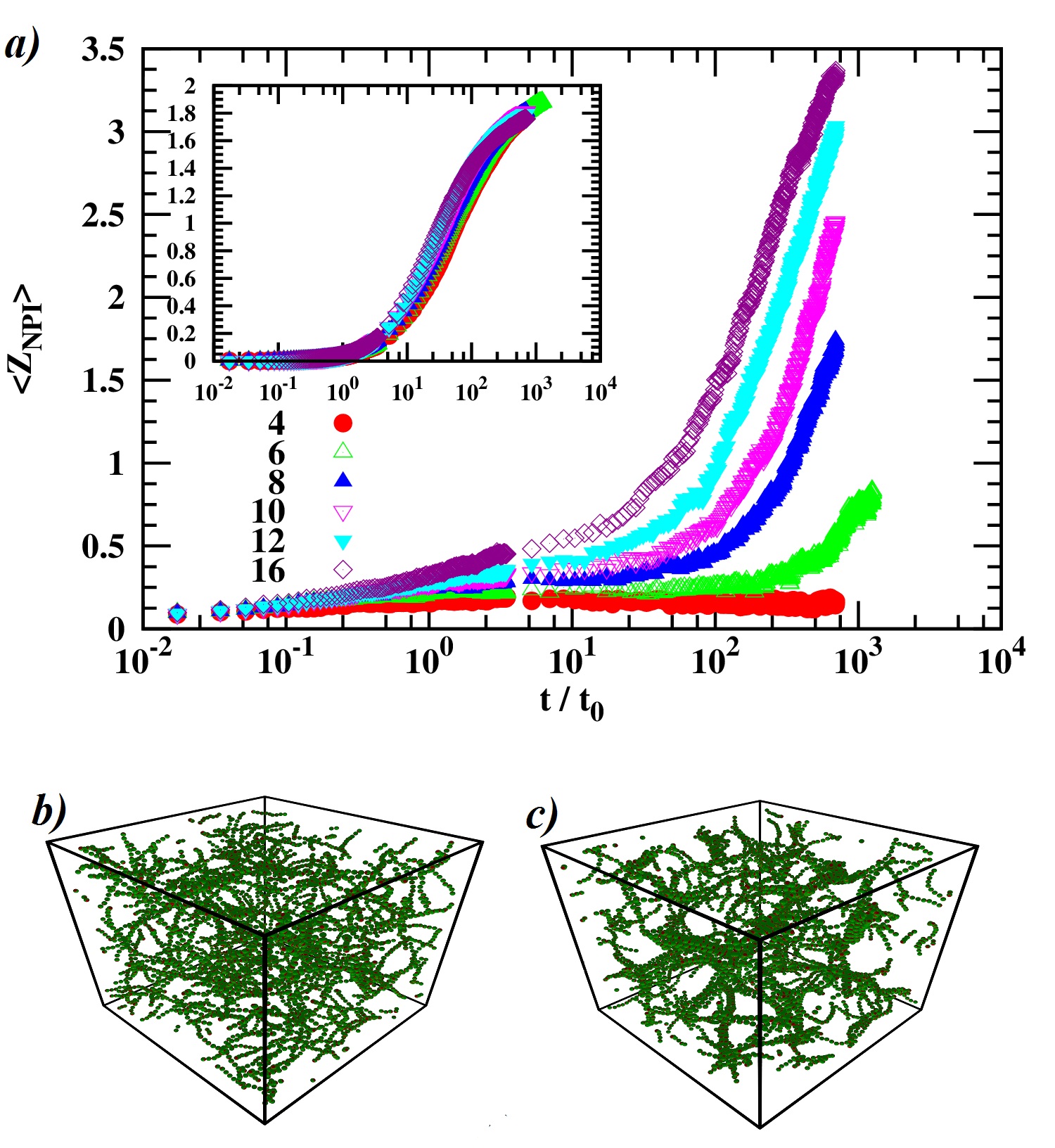}
  \caption{ The average bonded neighbors for $NPI$ type cluster $<$$Z_{NPI}$$>$ is plotted with respect to reduced time at different $B_{att}$ values as indicated in the figure for $\omega=22.5^{\circ}$. The inset shows average bonded neighbors for $P$ type cluster $<$$Z_P$$>$. (b). Visual images of chains formed at $B_{att}=4$.(c). Visual image of system formed at $B_{att}=6$, chains have started to transform into bundles.}
  \label{differentbatt}
\end{figure}

In order to understand the exact solvent condition which leads to the formation of bundles, we simulate the system at different $B_{att}$ values and follow their average number of neighbors $<$$Z_P$$>$ and $<$$Z_{NPI}$$>$.  In Fig. \ref{differentbatt}, we have plotted $<$$Z_{NPI}$$>$ and $<$$Z_P$$>$ in inset as a function of reduced time for $\omega=22.5^{\circ}$ and different $B_{att}$ values as indicated in the figure. We observe that $<$$Z_P$$>\sim2$ for all $B_{att}$ values indicating that each patch forms single bond for $\omega=22.5^{\circ}$  and system consists of elongated structures like chains and bundles. This is also shown in the snapshot of system formed at $B_{att}=4$ and $\omega=22.5^{\circ}$ at $t/t_0=1800$ in Fig. \ref{differentbatt}b. We observe that $<$$Z_{NPI}$$>$ accelerates for $B_{att}\geq6$ indicating the initialisation of chain to bundle transition as also shown in the snapshot of system formed at $B_{att}=6$ and $\omega=22.5^{\circ}$ at $t/t_0=1800$ in Fig. \ref{differentbatt}c. In this snapshot, we clearly observe that some of the chains have transformed into bundles as we move from $B_{att}=4$ to $B_{att}=6$ due to increase in the strength of isotropic interaction. This implies that chain to bundle transformation happens when quality of solvent deteriorates. In Fig. \ref{B12NPI}c, we show the snapshot of system formed at $\omega=22.5^{\circ}$, $B_{att}=12$ at $t/t_0=1800$ and we observe that all chains have transformed into bundles and system has percolated. In order to show the structural difference between bundles, chains and networks formed, we calculate the structural characteristics of different structures. 


\subsection{Structural Analysis}
\begin{figure}
\includegraphics[height=15cm,width=8.5cm]{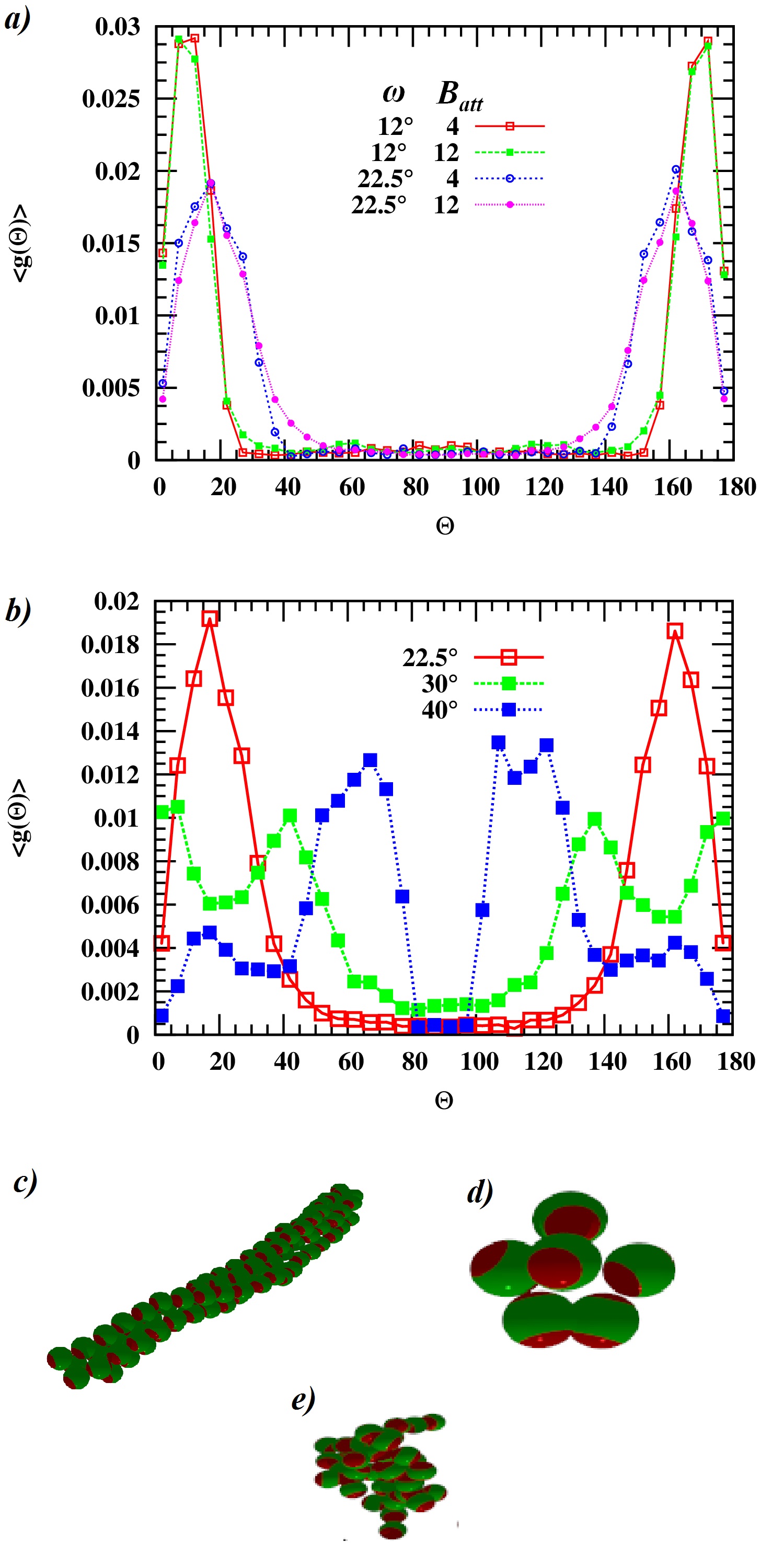}
  \caption{Distribution $g(\theta)$ of the angle $\theta$ $(cos^{-1}(\hat{\bf v}_{i}.\hat{\bf v}_{j}))$ for all pairs of nearest neighbors for different structures. (a). Open symbols indicate $g(\theta)$ values of system formed at $B_{att}=4$ and closed symbols indicate $g(\theta)$ values of system formed at $B_{att}=12$. Circles indicate systems formed at $\omega=22.5^{\circ}$ and squares indicate systems formed at $\omega=12^{\circ}$.(b). Structures formed at $B_{att}=12$ and different $\omega$ values as indicated in the figure. (c) Image of bundle formed at $B_{att}=12$ and $\omega=22.5^{\circ}$. (d). Image of a typical flower like structure formed at $B_{att}=12$ and  $\omega=30^{\circ}$. (e). Image of a typical agglomerate formed at $B_{att}=12$ and  $\omega=40^{\circ}$. }
  \label{gtheta3}
\end{figure}

$g(\theta)$ is the distribution of angle $\theta$ $=$ $(cos^{-1}(\hat{\bf v}_{i}.\hat{\bf v}_{j}))$ for all pairs of nearest neighbors. If two patch vectors are in same direction then $\theta=0^{\circ}$ and if they are facing each other, $\theta=180^{\circ}$. This distribution shows well defined peaks for any ordered structure and is flat for completely disordered structure\cite{sciortino2009phase,li2012model,sciortino2010numerical}. In Fig. \ref{gtheta3}a, we have plotted the $g(\theta)$ as a function of $\theta$ where circles indicate the $g
(\theta)$ values for systems formed at $\omega=22.5^{\circ}$ and squares indicate the $g(\theta)$ values for systems formed at $\omega=12^{\circ}$. The filled symbols indicate $g(\theta)$ values for systems formed at $B_{att}=12$ and open symbols indicate for $B_{att}=4$. As shown earlier, we observe the formation of chains and bundles at $B_{att}=4$ and $12$ respectively for $\omega\leq22.5^{\circ}$. The image of one such isolated bundle is shown in Fig. \ref{gtheta3}c, which clearly indicates that in all chains, patches of particles are facing each other and they are forming irreversible $P$ bond. Inter-chain bonds are formed due to isotropic potential and they are reversible. For chains and bundles formed at $\omega=22.5^{\circ}$, $g(\theta)$ shows clear double peaks at $20^{\circ}$ and $160^{\circ}$ due to $P$ bonds and additional tails for bundles at $40^{\circ}$ and $140^{\circ}$ due to $NPI$ bonds. For chains and bundles formed at $\omega=12^{\circ}$, $g(\theta)$ again shows double peaks which are shifted to $10^{\circ}$ and $170^{\circ}$ due to decreased patch size, but the formation of tails is similar to what we observed for $\omega=22.5^{\circ}$.  In Fig. \ref{gtheta3}b, we have plotted the $g(\theta)$ as a function of $\theta$ for $B_{att}=12$ and different $\omega$ values as indicated in the figure. For $\omega=30^{\circ}$, we observe networks consisting of many ordered structures which looks like flower arrangement as shown in the Fig. \ref{gtheta3}d. Due to this kind of arrangement, we observe four peaks at $0^{\circ}$, $45^{\circ}$, $135^{\circ}$ and $180^{\circ}$. Patch vectors of particles that are arranged in circle makes angles $45^{\circ}$ or $135^{\circ}$ with each other and the particle which is just below the central particle makes angle equals to $0^{\circ}$ or $180^{\circ}$ with the central particle. For $\omega=40^{\circ}$, we observe the formation of agglomerates as shown in the Fig. \ref{gtheta3}e and particles try to accommodate themselves in a way such that maximum number of patchy bonds are formed. Due to this, we observe the peaks that are shifted inwards at $70^{\circ}$ and $110^{\circ}$. Thus, the distribution function $g(\theta)$ can also be used to classify the different kinds of self assembled structures efficiently. 

\subsection{Influence of immobile obstacles on the formation of bundles }
\begin{figure}
\includegraphics[height=12cm,width=8.5cm]{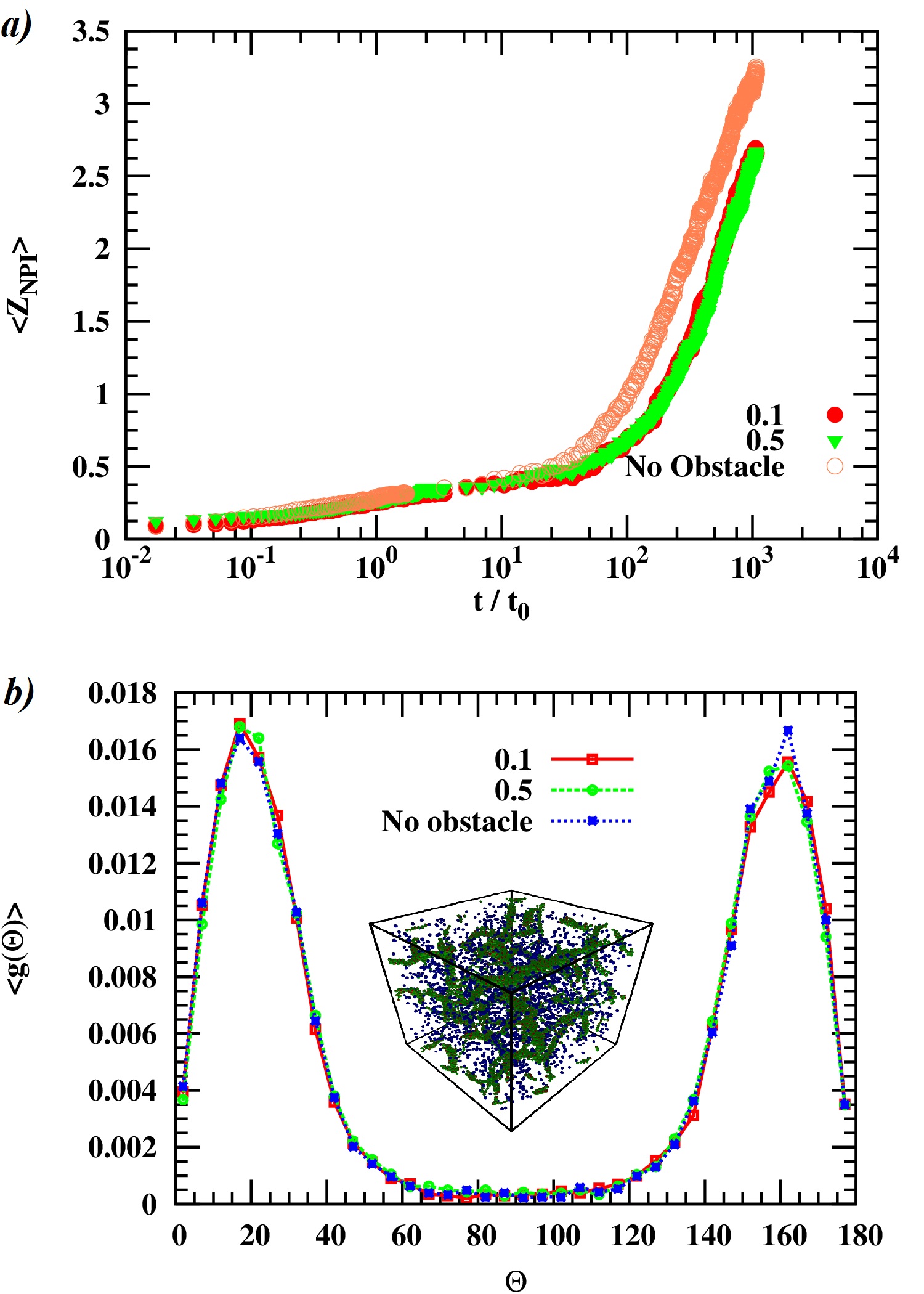}
  \caption{(a). $<$$Z_{NPI}$$>$  is plotted for $B$ particles with respect to reduced time at $\omega=22.5^{\circ}$ and $B_{att}=12$ in presence of $N_O$ immobile$(I)$ obstacles for a range of $C_O$ values as indicated in the figure. $<Z_{P}>$ is plotted for $B$ particles as a function of physical time in the inset. (b). Distribution $g(\theta)$ of the angle $\theta$ for all pairs of nearest neighbors of $B$ particles. The snapshot of the system formed at $t/t_0=1800$ at $C_O=0.5$ is shown as an inset of Fig. \ref{immobilegraph}b, here $I$ particles are indicated by blue spheres and $B$ particles are indicated by green spheres. We can observe the formation of bundles of $B$ particles in presence of $I$ obstacles.   }
  \label{immobilegraph}
\end{figure}

In Fig. \ref{immobilegraph}a, we plot $<$$Z_{NPI}$$>$ as a function of reduced time for $B$ particles at $B_{att}=12$ and $\omega=22.5^{\circ}$ in presence of $N_O$ immobile($I$) obstacles whose fraction $C_O$ is indicated in the figure. We observe that $<$$Z_{NPI}$$>$ is less for system which aggregates in presence of $I$ particles and this implies that kinetics of formation of bundles has slowed down in presence of immobile obstacles. This slowing down of aggregation kinetics is attributed to the hindrance provided by immobile obstacles and excluded volume effect. It takes time for clusters of $B$ particles to move around these obstacles. The kinetics of aggregation is almost independent of the fraction of obstacles, this indicates that even in presence of very low fraction of immobile obstacles, the formation of bundles is slowed down. In Fig. \ref{immobilegraph}b, we have plotted the $g(\theta)$ as a function of $\theta$ for $B$ particles and the fraction of $I$ particles is as indicated in the figure. We observe that peaks in presence of $I$ particles overlaps with the peaks when no such obstacles are present. It indicates that bundles are formed, only the kinetics of bundle formation has slowed down. The snapshot of the system formed at $t/t_0=1800$ at $C_O=0.5$ is shown as an inset of Fig. \ref{immobilegraph}b, which clearly indicates the formation of bundles of $B$ particles in presence of $I$ obstacles.  

\begin{figure}
\includegraphics[height=12cm,width=8.5cm]{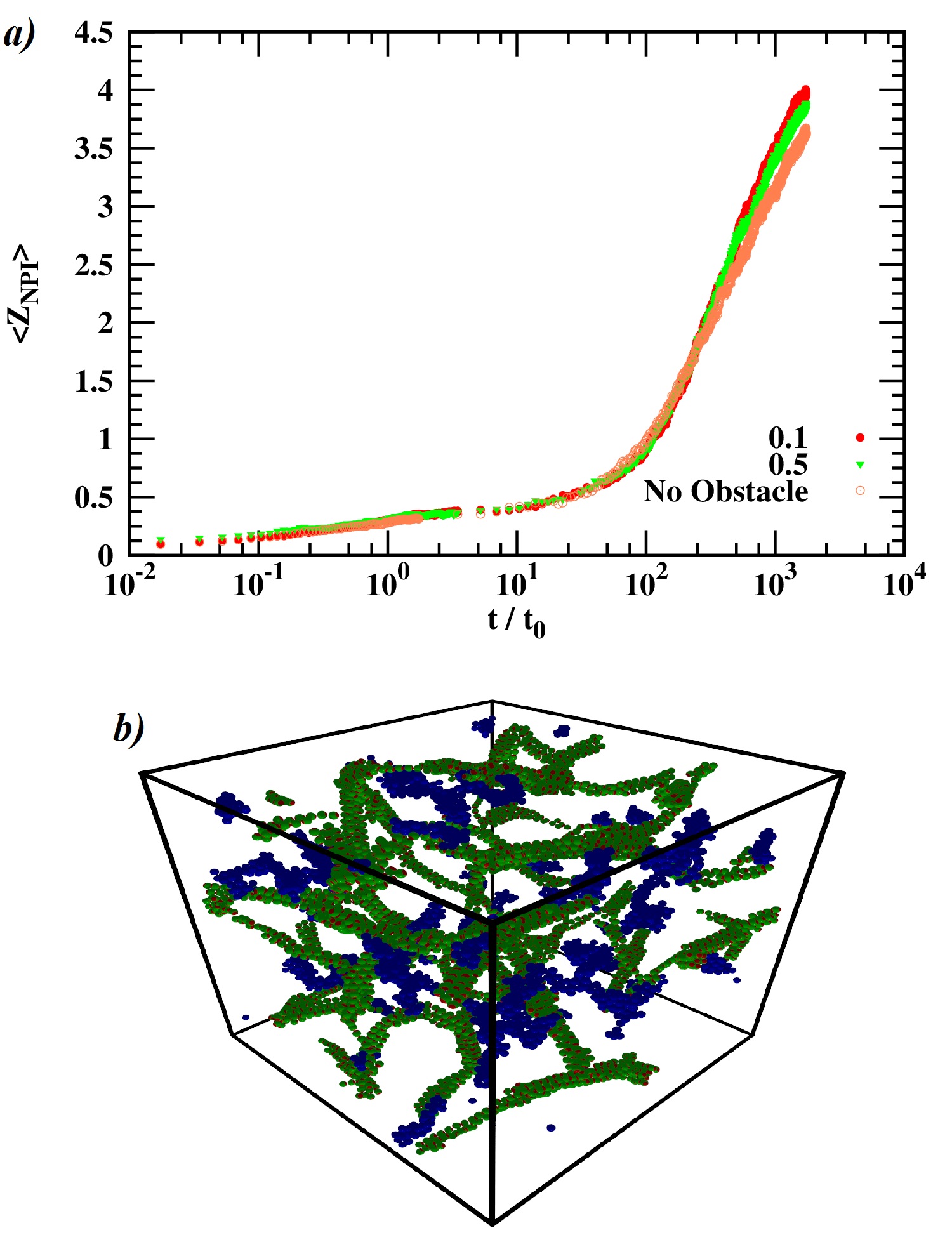}
  \caption{(a). $<$$Z_{NPI}$$>$  is plotted for $B$ particles with respect to reduced time at $\omega=22.5^{\circ}$ and $B_{att}=12$ in presence of $N_O$ mobile($M$) obstacles for a range of $C_O$ values as indicated in the figure.(b). The snapshot of the system formed at $t/t_0=1800$ at $C_O=0.5$ is shown, here $M$ particles are indicated by blue spheres and $B$ particles are indicated by green spheres. We can observe the formation of bundles of $B$ particles in presence of network of $M$ obstacles.}
  \label{obstacle4590}
\end{figure}
\begin{figure}
\includegraphics[height=12cm,width=8.5cm]{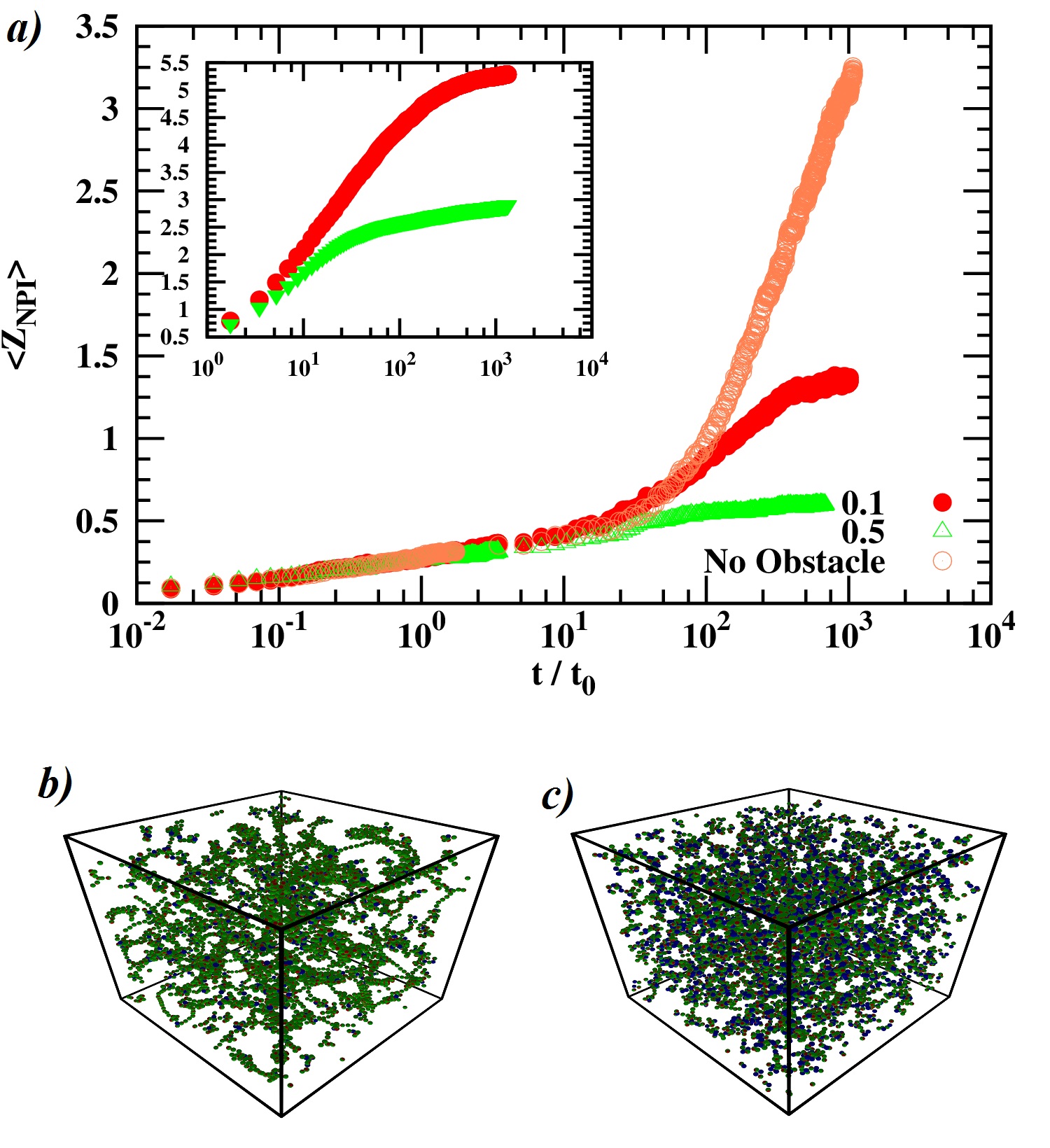}
  \caption{(a). $<$$Z_{NPI}$$>$  is plotted for $B$ particles with respect to reduced time at $\omega=22.5^{\circ}$ and $B_{att}=12$ in presence of $N_O$ mobile($M$) obstacles of patch size $\omega=90^{\circ}$ for a range of $C_O$ values as indicated in the figure. (b). The snapshot of the system formed at $t/t_0=1800$ at $C_O=0.1$ is shown, here $M$ particles are indicated by blue spheres and $B$ particles are indicated by green spheres. We can observe the agglomerates formed by the $B$ particles with $M$ obstacles and formation of bundles is partially inhibited. (c). The snapshot of the system formed at $t/t_0=1800$ at $C_O=0.5$ is shown, we observe the agglomerates formed by the $B$ particles with $M$ obstacles.  }
  \label{22545}
\end{figure}

\subsection{Influence of mobile patchy obstacles with inter-particle attraction and no intra-particle hard sphere repulsion on the formation of bundles }
\begin{figure}
\includegraphics[height=6cm,width=8.5cm]{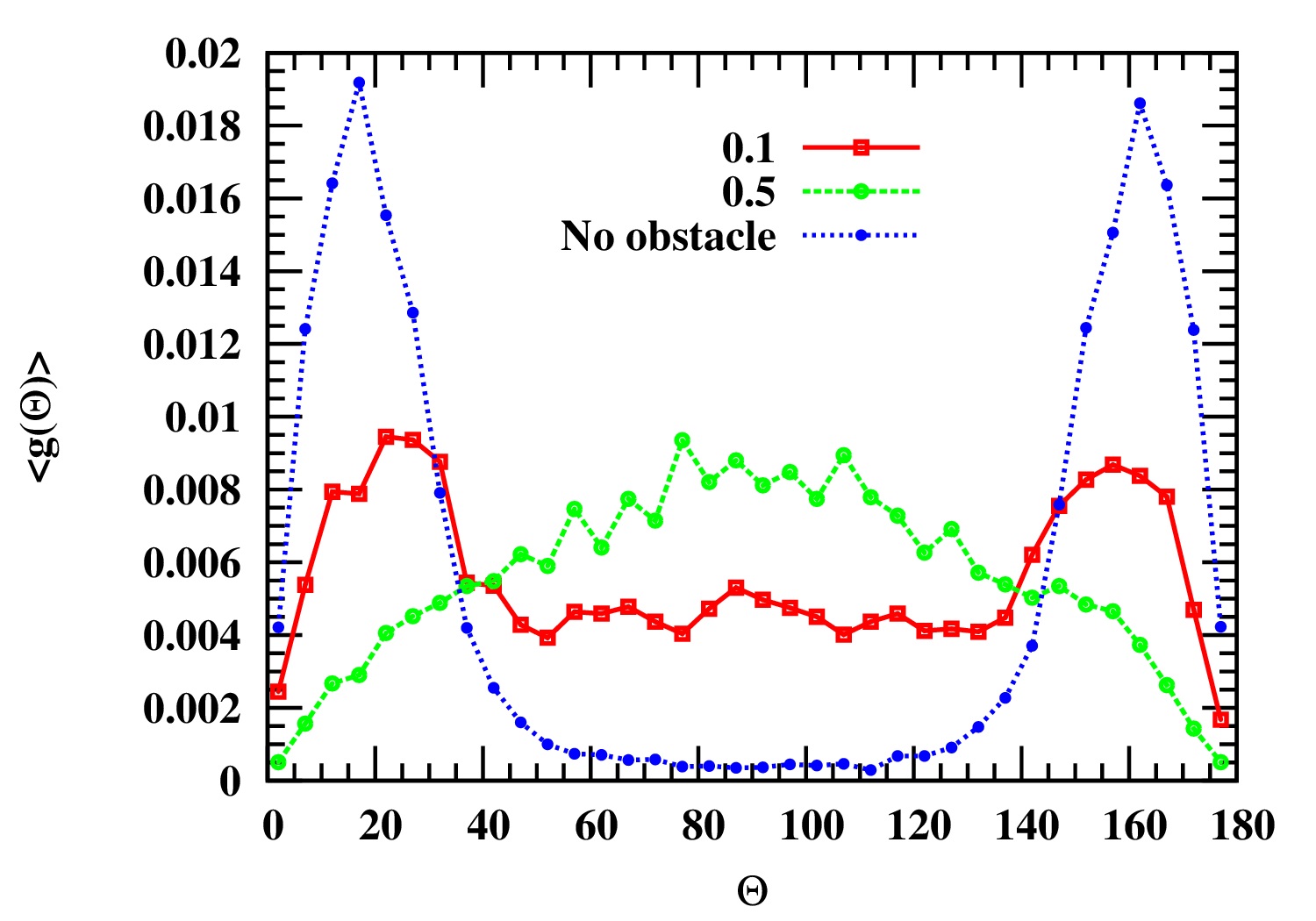}
  \caption{Distribution $g(\theta)$ of the angle $\theta$ for all pairs of adjacent $B$ particles at different $C_O$ values as indicated. }
  \label{gtheta8}
\end{figure}

In Fig. \ref{obstacle4590}, we  plot $<$$Z_{NPI}$$>$ as a function of reduced time for $B$ particles at $B_{att}=12$ and $\omega=22.5^{\circ}$ in presence of $N_O$ mobile $(M)$ obstacles and fraction of these obstacles $C_O$ is indicated in the figure. $M$ particles interact via  irreversible isotropic square well interaction with each other and via hard sphere repulsion with $B$ particles.  We observe that kinetics of formation of bundles has slightly accelerated in presence of $M$ obstacles. It is indicated by slightly higher $<$$Z_{NPI}$$>$ value in presence of $M$ particles as compared to $<$$Z_{NPI}$$>$ when no such obstacles are present. Here $M$ particles interact via irreversible isotropic interaction and thus it leads to the formation of network of obstacles. Here, the excluded volume effect has reduced due to the formation of clusters of $M$ particles as compared to the case of immobile $I$ particles and this leads to the slight acceleration in the kinetics of formation of bundles\cite{babu2008tracer}. The snapshot of the system formed at $t/t_0=1800$ at $C_O=0.5$ is shown in Fig. \ref{obstacle4590}b, here $M$ particles are indicated by blue spheres and $B$ particles are indicated by green spheres. This clearly shows the formation of bundles of $B$ particles in presence of network of $M$ obstacles.

\subsection{Influence of mobile patchy obstacles with intra-particle interaction and no inter-particle interaction on the formation of bundles }
In Fig. \ref{22545}, we  plot $<$$Z_{NPI}$$>$ as a function of reduced time for $B$ particles at $B_{att}=12$ and $\omega=22.5^{\circ}$ in presence of $N_O$ mobile$(M)$ patchy obstacles of patch size $\omega=90^{\circ}$ and fraction of these obstacles $C_O$ is indicated in the figure. $M$ particles interact via hard sphere repulsions with each other, whereas they interact via potential given by Eq. \ref{e.1} with $B$ particles. We observe that formation of bundles has been inhibited in this case indicated by low value of $<$$Z_{NPI}$$>$ in Fig. \ref{22545}. As the fraction of $M$ particles increases, bundles ceases to exist as shown by very low value of $<$$Z_{NPI}$$>$ at $C_0=0.5$. In inset, we  plot $<$$Z_{P}$$>$ as a function of reduced time for $M$ particles. We observe that $<$$Z_P$$>$ is higher for $C_O=0.1$ than $C_O=0.5$, this is due to the fact that if we have large number of $M$ particles, they form agglomerates with $B$ particles such that valencies of $B$ particles get exhausted. These agglomerates ceases to grow as $M$ particles interact via hard sphere interactions with each other and $<$$Z_P$$>$ of $M$ particle stagnates. These agglomerates inhibits the longitudinal growth of $B$ particles into chains and hence bundles.  
The visual image of the agglomerates formed of $B$ and $M$ particles along with chains of $B$ particles is shown at $C_0=0.1$ in Fig. \ref{22545}b. In  Fig. \ref{22545}c, we show the visual image of system consisting of $B$ and $M$ particles at $C_0=0.5$. In this case, we observe that formation of chains and bundles is almost inhibited and only agglomerates of $B$ and $M$ particles are formed. In order to show the effect of these $M$ particles on the formation of bundles of $B$ particles, we calculate $g(\theta)$ of $B$ particles in presence of $C_0$ fraction of $M$ particles as indicated in the Fig. \ref{gtheta8}. We observe that in presence of small fraction of $M$ particles, the formation of bundles is partially inhibited indicated by reduced peaks at $20^{\circ}$ and $160^{\circ}$ and finite probability of intermediate angles. As $C_0$ is increased to $0.5$, we observe that all angles have finite probability and this is a clear indication of complete inhibition of formation of chains and bundles. This implies that addition of small amount of $M$ particles can inhibit partially/completely the bundle formation. 
\section{\label{sec:discussion}Discussions and Conclusions}
In the present work, the role of solvent condition and patch size is investigated for two-patch colloidal particles. Different structures such as chains, bundles, flower like structure and agglomerates with distinctive structural characteristics are achieved by tuning the patch size and solvent conditions. It is important to understand the aggregation of bundle formation in presence of inhibitors to develop and evaluate the drugs\cite{ribarivc2018peptides}.  Various therapies are developed to control the formation of amyloid fibres such as antioxidants\cite{zandi2004reduced} and anti-inflammatory agents\cite{aisen2006phase,scharf1999double}, but they are not effective. The major concern is to develop molecules that inhibit the formation of amyloid fibres without disrupting the normal biochemical processes in the body. Many inhibitors are studied which attenuate the formation of amyloid fibres. They form covalent or non-covalent bonds with the aggregation products in a nonspecific manner\cite{feng2008small,lendel2010detergent}. We have investigated the kinetics of aggregation of bundles in presence $M$ and $I$ particles. We have shown that the presence of $I$ particles slows down the kinetics of formation of bundles but does not inhibit it. The presence of $M$ particles with inter-particle attraction and intra-particle hard sphere repulsion slightly accelerates the formation of bundles. We have also shown that in presence of $M$ particles with intra-particle attraction and inter-particle hard sphere repulsion, the formation of bundles is inhibited significantly even in presence of small number of $M$ particles.
 We know that the first step in the formation of amyloid fibrils is the aggregation of colloidal spheres into linear chains \cite{wasmer2008amyloid}. In the present study, we are showing that chains form due to the presence of directional interaction. If the valency of these patches are exhausted by the presence of other molecules or particles ($M$ particles), the formation of linear chains and hence bundles can be inhibited. 

Experimentally, it has been found that vitamin K3 inhibits the fibril formation of Lysozyme protein\cite{alam2016vitamin} by forming smaller sized aggregates that are less toxic. In our work also, we observe the formation of smaller aggregates that are distributed in box in presence of $M$ particles. $M$  particles surround $B$ particles thereby completely exhausting their valencies. Siddiqi \textit{et al.}\cite{siddiqi2018elucidating} studied the kinetics of formation of insulin fibres and observed that kinetic curve possesses a lag phase and after a certain time it increases in an exponential manner and a similar kinetic behavior is observed in this work, see Fig. \ref{B12NPI}a. In presence of peptide P4, the kinetics of formation of fibres is decreased by a factor of $\sim5.6$\cite{siddiqi2018elucidating}. In our work in presence of  $C_O=0.5$ fraction of $M$ obstacles, $<Z_{NPI}>$ decreases by a factor of $5.36$ at $B_{att}=12$ and $t/t_0=1800$. They also suggested that peptides probably masked all available hydrophobic patches. We are providing a model showing how inhibitors interact and inhibits the bundle formation. Our work will give way for discovery of such particles that can inhibit the amyloid formation by aggregating with the bundle forming particles and exhausting their patchy valencies. Upto now, all inhibitors are forming non specific bonds with the bundle forming particles. But in this work we have predicted that in order to form chains, particles need to have patchy interaction and if these patchy valencies can be exhausted, bundle formation can be inhibited completely.

\section{Acknowledgements}
We thank HPC padum and Badal of IIT Delhi for providing us the necessary computational resources.

\end{document}